\documentclass[onecolumn,superscriptaddress,amsmath,amssymb,aps,prfluids]{revtex4-2}

\usepackage{microtype}

\usepackage{graphicx}
\usepackage{dcolumn}
\usepackage{bm}
\usepackage{caption}
\usepackage{xcolor}

\begin{document}

\preprint{APS/123-QED}

\title{Pattern Formation in Bioconvection of \textit{Thiovulum} in a Hele-Shaw Chamber}

\author{Trey Johnson}
\author{George A. Schaible}
\author{Yujia Qi}
\author{Fridtjof Brauns}
\affiliation{Department of Mechanical Engineering, University of California, Santa Barbara, California 93106, USA}

\author{Alexander Cohen}
\affiliation{Department of Mathematics, Massachusetts Institute of Technology, Cambridge, Massachusetts 02139, USA}

\author{Jean-Marie Volland}
\email{jvolland@ucsb.edu}
\affiliation{Department of Molecular, Cellular, and Developmental Biology, University of California, Santa Barbara, California 93106, USA}

\author{Ousmane Kodio}
\email{kodio@ucsb.edu}
\affiliation{Department of Mechanical Engineering, University of California, Santa Barbara, California 93106, USA}

\date{\today}

\begin{abstract}
This paper is associated with a video winner of a 2025 American Physical Society's Division of Fluid Dynamics (DFD) Gallery of Fluid Motion Award for work presented at the DFD Gallery of Fluid Motion. The original video is available online at the Gallery of Fluid Motion, \url{https://doi.org/10.1103/APS.DFD.2025.GFM.V045}
\end{abstract}

\maketitle

We investigate bioconvection in a colony of \textit{Thiovulum} sp. ST bacteria, a recently isolated enrichment culture, confined within a Hele-Shaw chamber. Driven by chemotactic and gravitactic responses, the cells collectively develop striking emergent patterns and convection-like dynamics. Starting from a dense, homogeneous suspension, the swimming bacteria generate large-scale bioconvective flows within minutes. Although these flows resemble thermal convection, they arise in the absence of an imposed temperature gradient. Instead, they arise from the collective swimming of bacteria responding to oxygen gradients and gravity.

Initially, the cells form a dense layer near the water surface after which spatially periodic density fluctuations along the layer then start to develop. Subsequently, regions of high cell density spontaneously develop into ``plumes" of downward-swimming cells sustained by collective cell motion. The plume pattern exhibits a characteristic wavelength that appears to depend on the experimental conditions. After 30--50 minutes following homogenization, the plumes mature into stable structures interspersed with low-density ``clouds" of upward-swimming cells (see Fig.~1). These patterns persist as long as the bacteria remain motile, ultimately limited by the availability of an energy source, namely, hydrogen sulfide and oxygen dissolved in the water. In our experiments, the patterns remain observable for up to 4 hours, corresponding to the duration of our longest measurements in the Hele-Shaw chamber.

\begin{figure}[htbp]
\centering
\includegraphics[width=\textwidth, trim=0pt 0pt 0 0pt, clip]{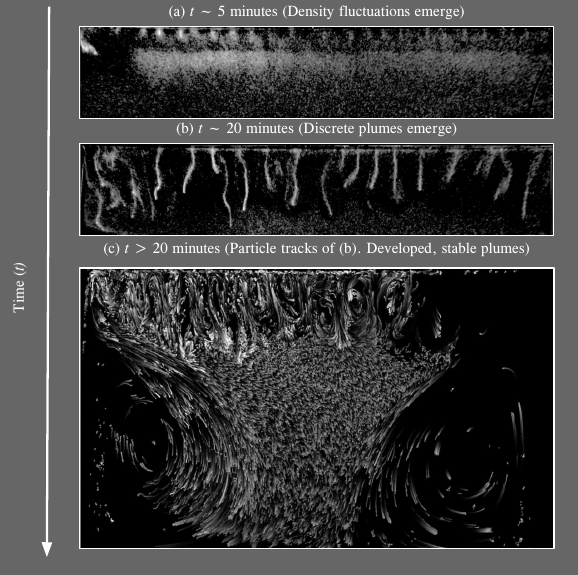}%
\caption{\label{fig:plume_progression} \textbf{Plume Progression.} 
(a) Cropped, contrast-processed video frame about 5 minutes after homogenization: density-fluctuation patterns form near the air-liquid interface. (b) About 20 minutes after homogenization: visible plumes emerge at regular spatial intervals. These plumes' fingers extend further in time and are very unstable, forming, diffusing, and merging over the span of a few minutes. (c) The system reaches a stable, spatially periodic plume structure, with low-density clouds of rising cells interspersed between the plumes. These structures remained stable for as long as our longest observation trial of 4 hours. This image is a view of particle tracks showing the small-scale bioconvection patterns near the surface and large-scale patterns near the corners. Note: The lens geometry induces a pincushion distortion, stretching the image near the corners.}
\end{figure}

To resolve and track individual cells while retaining a large field of view, we developed a high-contrast imaging setup, shown in Fig.~2a, which mimics dark-field microscopy by preventing unscattered light from reaching the camera. Cells of \textit{Thiovulum} sp. ST are well suited to this imaging approach: they are amongst the largest known free-swimming bacteria, with cell diameters of $11\mu\mathrm{m}$ and amongst the fastest, reaching local swimming speeds of up to $1\,\mathrm{mm\,s^{-1}}$ in our experiments. In our observations, we also noticed that \textit{Thiovulum} sp. ST may exhibit a response to light; we therefore illuminated the chamber uniformly to avoid introducing local inhomogeneities. We used a Hele-Shaw chamber ($75\text{ mm} \times 50\text{ mm} \times 1\text{ mm}$) to confine the motion to approximately two dimensions, making it much easier to image the collective patterns and simplifying mathematical modeling and analysis \cite{petroff14}. The bacteria were cultivated in a salt-water medium, mimicking their natural environment, injected into the chamber through its exposed surface, and homogenized using a small wire. Sodium dithionite was added to steepen the oxygen gradient and thereby quickly promote bioconvective flows \cite{wirsen78}. We recorded time-lapse video at one second intervals for several hours. The frames were then normalized and filtered to isolate cell-sized features, enabling individual-cell displacements to be tracked using a least-squares method (Fig.~2d).

The behavior of \textit{Thiovulum} sp. ST is rooted in its sulfur-oxidizing metabolism. Like most sulfur-oxidizing bacteria, the cells store sulfur in intracellular granules and are limited by the supply of oxygen \cite{jorgensen83}; consequently, chemotaxis draws them toward regions of higher oxygen concentrations \cite{thar01}. Gravitaxis acts alongside this chemotactic response: the cells are negatively buoyant and, owing to their internal sulfur stores, possess an offset of the center of mass from the center of buoyancy, exerting a torque that gradually aligns the swimming direction with the direction of gravity \cite{thar01}. In addition, shear flows exert torques leading to so-called gyrotaxis shaping the narrow plumes \cite{bees20}.

\begin{figure}[htbp]
\includegraphics[width=\textwidth, trim=0pt 0pt 0pt 0pt, clip]{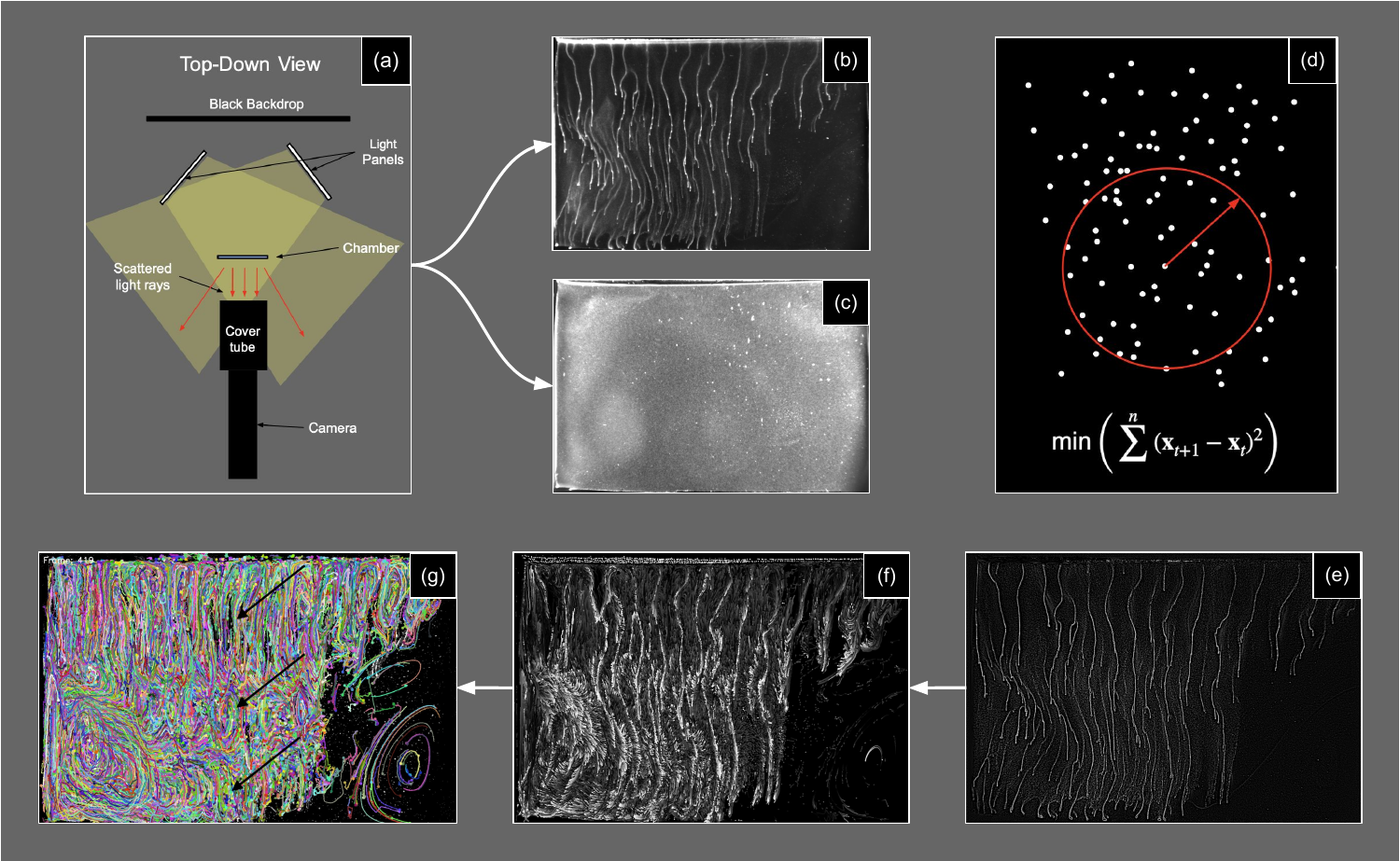}%
\caption{\label{fig:video_processing} \textbf{Video-processing pipeline.} 
(a) Imaging setup: a CCD camera with a macro zoom lens images the Hele-Shaw chamber. Two light panels are angled behind the chamber, with a black backdrop placed further behind. A black tube extends past the lens, blocking unscattered source light from the focal path to create a high-contrast, dark-field-like image. (b) Raw video frame of the pattern formation in the Hele-Shaw chamber. (c) Example of background image acquired immediately after homogenization. Ten such images were median-stacked into a single image background frame capturing lighting irregularities, dust in the optical path, and scratches on the chamber glass. (d)~Small features are identified in a frame processed with a short-wavelength bandpass filter. A least-squares displacement tracking algorithm links features between consecutive frames, choosing the matches that minimize the total sum-of-squares displacement between the positions of particles in consecutive frames $x_{t+1}$ and $x_{t}$. (e) Each raw video frame such as that shown in Fig.~2b is divided by the median background frame shown in Fig.~2c to normalize the image and remove artifacts caused by dust and scratches. A nonlinear intensity stretch increases the visibility of faint cells. (f) Example frame showing the resulting particle tracks. (g) Long particle tracks, randomly colored for illustration. Black arrows mark the vertical levels of bioconvection vortices. Near the boundaries between these levels, cells move mostly horizontally.}
\vspace{-10pt}
\end{figure}

The chemotactic response is observable in the low-density clouds of cells swimming upward toward the exposed air-liquid surface, where oxygen diffuses into the medium. As cells accumulate near the surface, they create a density inversion that likely triggers a Rayleigh-Taylor-type instability, producing the downward flows that initiate plume formation \cite{plesset74}. Near the boundary of the high-density regions we observe the formation of a “veil” of cells swimming parallel to the boundary, their direction of motion aligned with the chemical gradient. After descending within a plume for some time, the cells disperse from it and resume their upward chemotactic and gravitactic swimming. The result is a sustained, self-organized pattern that resembles vortices in thermal convection \cite{bees20}. These patterns occur on multiple scales (Fig.~1c): small-scale vortices form around the plumes near the surface, while large-scale vortices occupy the remainder of the chamber. Because the large-scale circulation consistently appears in the presence of chemotactically driven bacteria, we infer that it, too, is a consequence of the swimming-induced flows. Surprisingly, while mature plumes typically extend the entire length of the chamber, the bioconvection-induced vortices do not (Fig.~2g). Instead, distinct vertical levels of bioconvection vortices form. What dictates how and where a plume ends and how cells leave a plume remains unclear.

\enlargethispage{\baselineskip}
Bioconvection in \textit{Thiovulum} adds to the rich variety of pattern-forming bacterial systems \cite{bees20}. Combining high-contrast imaging with computational image processing, we isolated and tracked individual cells at high resolution across a large field of view, revealing how their collective motion drives complex macroscopic fluid transport. Collective flows generated by similar bacteria have been shown to enhance the efficiency of oxygen transport \cite{petroff14}. Unlike prior work done on tethering \textit{Thiovulum majus} \cite{petroff14}, we observe the cells of \textit{Thiovulum} sp. ST swimming parallel to strong oxygen gradients while forming these patterns. This behavior may optimize their simultaneous access to oxygen-rich and sulfide-rich water. Cells of \textit{Thiovulum} sp. ST are usually large and contain intracellular sulfur granules that make them appear bright white under incident illumination, rendering their collective structures readily visible to the unaided eye. These features make them a particularly suitable biological system for studying collective behavior analogous to bird flocking or fish schooling \cite{Vicsek2012}. Thus, this system provides an accessible model for investigating self-organization across a broad range of physical and biological active matter systems. Further work is needed to determine how pattern features depend on the interplay of gravitaxis, chemotaxis, gyrotaxis, and cell density. The resulting quantitative measurements of the emergent patterns under varying conditions could provide direct tests for active matter theories.

\textit{Data availability:} The data that support the findings of this article are not publicly available. The
data are available from the authors upon reasonable request.
 
\vspace{-15pt} 
\bibliography{main}

\end{document}